\documentstyle{article}

\def\a{\alpha}
\def\b{\beta}
\def\g{\gamma}
\def\d{\delta}
\def\e{\epsilon}

\def\th{\theta}

\newcommand\bg{\begin{eqnarray}}
\newcommand\ed{\end{eqnarray}}
\newcommand\bgn{\begin{eqnarray*}}
\newcommand\edn{\end{eqnarray*}}

\def\D{\partial}
\def\ra{\Longrightarrow}
\def\rat{\rightarrow}
\def\disp{{\displaystyle}}

\def\Xr{X^{R}}
\def\Xxr{\Xr_{x}}
\def\Xvr{\Xr_{v}}
\def\Xtr{\Xr_{t}}

\def\Xtl{{X_{t}}^{L}}

\def\xt{\tilde{X}}

\def\ET{{{\xt}_{t}}^{R}}
\def\XT{{{\xt}_{x}}^{R}}
\def\KT{{{\xt}_{v}}^{R}}

\def\ETl{{{\xt}_{t}}^{L}}
\def\XTl{{{\xt}_{x}}^{L}}
\def\KTl{{{\xt}_{v}}^{L}}

\def\tec{\tilde{\th}}

\def\L{\cal{L}}

\begin{document}
\vskip.15in
\rightline{BROWN-HET-921}
\rightline{ August 1993}
\vskip .5in
\centerline{\LARGE{\bf{Hidden invariance of the free classical particle}}}
\leftline{}
\leftline{}
\centerline{\large{Santiago Garc\' \i a}}
\leftline{}
\smallskip
\centerline{\large{Department of Physics}}
\smallskip
\centerline{\large{Brown University}}
\smallskip
\centerline{\large{Providence~~~ RI~02912}}
\leftline{}
\leftline{}
\leftline{}
\leftline{}
\leftline{}
\begin{abstract}
A formalism describing the dynamics of classical and quantum systems
from a group theoretical point of view is presented.
We apply it to the simple example of the classical free particle.
The Galileo group $G$ is the symmetry group of the free equations
of motion. Consideration of the free particle Lagrangian semi-invariance
under $G$ leads to a larger symmetry group, which is a central extension
of the Galileo group by the real numbers. We study the dynamics associated
with this group, and characterize quantities like Noether invariants and
evolution equations in terms of group geometric objects. An extension of
the Galileo group by $U(1)$ leads to quantum mechanics.
\end{abstract}
\leftline{}
\leftline{}
\leftline{}
\leftline{}
\newpage
\section{Introduction}
\par
\leftline{}
The importance of symmetry in physics can hardly be
stressed \cite{ajp2}.
We present a formalism that exploits symmetry in order to
describe
certain classical and quantum systems.
The fundamental object in our approach is a Lie group
that represents the physical system under consideration.
All dynamical quantities (Noether invariants, evolution equations
$\ldots$)
are canonically defined in terms of geometric
quantities.

This formalism is akin in language
to Geometric Quantization \cite{geom},
whose goal is to build a quantum theory
from a classical one: a Hilbert space where quantum operators
reproduce the classical algebra of the
Poisson brackets.
But there is a crucial difference in philosophy. We
do not assume the existence, a priori, of any classical system:
quantum or classical nature are determined, as we shall see
in this work, by the geometric structure
of the group {\cite{aldaya}}.

We haven chosen the free particle as a simple example
to present our ideas.
Since our aim is to write an
introductory review, we do calculations in detail, using
coordinates rather than more elegant and concise notation.
Quantities
and concepts are introduced gradually.
Although sometimes this may seem very ad-hoc,
we think that readers
not familiar with differential geometry or group theory will
be able to follow our reasonings.

First, we define the
Galileo group {\cite{ajp1}}.
Although the equations of motion
of the free particle are Galilean invariant, it is not possible
to construct a Lagrangian strictly invariant under
Galilean transformations. We introduce then the
concept of semi-invariance.

The equivalence of the Lagrangians
${\cal{L}}$ and ${\cal{L'}}=
{\cal{L}}+ dF/dt $, where $F$ is an arbitrary
function of coordinates and time,
leads to the postulate of a new symmetry group. It is remarkable
that the infinitesimal transformations of this group can be found
by considering the action of the Galileo group on the
arbitrary function $F(x,t)$. In an abuse of language,
we have called this symmetry {\em{hidden}} because it is not
a symmetry involving the dynamical variables of the particle,
coordinate and velocity. Nevertheless, it is related to its
mass in a subtle way.

The next section is devoted to the study of the new symmetry group.
We find that this group
is a central extension of the Galileo group by the real numbers.
Central extensions have a natural fiber bundle structure.
We introduce
the canonical form $\Theta$, a connection in the bundle.
Symmetries of $\, \Theta \, $ define Noether
invariants and the classical equations of motion for the particle.
Finally, the  Hamilton-Jacobi equation is obtained.
Symplectic structure and Poisson brackets
are discussed in appendix II.

Historically (\cite{bargmann}, \cite{Wigner}),
the phase invariance of the scalar product for two
quantum mechanical wave functions
$$
\langle \psi \,\vert \,\chi \rangle =
\int{\,d x \,{\psi^{*} (x)} \,\chi (x) }
$$
under a $U(1)$ transformation
$$\psi \mapsto \zeta \psi \qquad , \qquad
\chi \mapsto \zeta \chi \qquad  \qquad / \quad \zeta \in U(1) $$
led to the consideration of {\it{ray}} representations of
Lie groups: representations defined up to
a phase.
This is the motivation for the last section of this paper,
where
we extend the Galileo group by  $U(1)$.
The Schr\"odinger equation
and the Heisenberg commutation relations are obtained.
We construct the
Hilbert space of wave functions.
The quantum operators (position, momentum)
acting on this Hilbert space are given
in terms of the right invariant
generators of the group.
Finally, we discuss the correspondence between the
classical and quantum theories.

The reader familiar with Geometric Quantization will recognize
all its basic ingredients
($\Theta$ is the quantization form in a contact manifold, $d \Theta$
is a presymplectic form, etc.\cite{aldaya}).
Nevertheless, this group quantization
formalism, besides providing a canonical definition of dynamical
quantities
based solely on group invariance, solves one essential difficulty in
Geometric Quantization: the
characterization of a polarization (constraint) that reduces
the representation of the symmetry group in the physical
Hilbert space . Since a group structure is present, a
polarization can be defined by considering a maximal
subalgebra of group generators, as shown in sections $6$
and $7$.

There are two appendices at the end of this paper. The first one
presents basic concepts of differential geometry (vector fields, forms,
integral curves ...), while the second deals with the
definition of the Poisson brackets (symplectic structure)
and the formulation of Noether's theorem in this group theoretical
language.

\newpage

\section {The Galileo Group}
Newton's equations of motion for a free particle of mass $m$
\bg
m { d^{2} {\bf{x}} \over d t^2}= 0
\label{gal0}
\ed
can be decomposed into two first order equations
\bg
{d {\bf{x}} \over d t}= {\bf{v}} \qquad \qquad m \,{ d {\bf{v}} \over d t}=
{\bf{0}}
\label{gal1}
\ed
where the position
${\bf{x}}$ and the velocity ${\bf{v}}$ are three-dimensional real vectors.
Equations (\ref{gal1}) are invariant under the following
{\em{Galilean}} transformations (\cite{aldaya}, \cite{ajp1},
\cite{bargmann}, \cite{Wigner})
\bg
{\bf{x}} \rat & {\bf{x}}'= & {\bf{R}}{\bf{x}} + {\bf{x}}_{0} +{\bf{v}}_{0} t
 \nonumber \\
{\bf{v}} \rat & {\bf{v}}'= & {\bf{R}}{\bf{v}} + {\bf{v}}_{0}
 \nonumber \\
t \rat & t' = & t + t_{0}
\label{gal11}
\ed
where $({\bf{x}}_{0},{\bf{v}}_{0},t_{0})$ are
constants and ${\bf{R}}$ is a rotation matrix, ${\bf{R}} \in SO(3)$.
Rotations complicate the notation unnecessarily
without playing any role in our discussion, so in the following
only Galilean transformations in one dimension, the line defined by the motion
of the particle, will be considered.
\bg
x \rat & x'= & x + {x}_{0} +{v}_{0} t
 \nonumber \\
v \rat & v'= & v + {v}_{0}
 \nonumber \\
t \rat & t' = & t + t_{0}
\label{gal2}
\ed
We can write (\ref{gal2})
in matrix form by introducing a four component column vector \cite{explan}
with an extra $1$ as the fourth component.
\bg
 \left( {\matrix{
x'\cr
v'\cr
t'\cr
1\cr}} \right) =
 \left( {\matrix{
1&0&v_{0}&x_{0}\cr
0&1&0&v_{0}\cr
0&0&1&t_{0}\cr
0&0&0&1 \cr} }  \right)
 \left( {\matrix{
x\cr
v\cr
t\cr
1\cr} } \right)
\label{matrix1}
\ed
The Galilean transformations form a group, the Galileo group $G$:
the composition law for the group parameters
$(x_{0},v_{0},t_{0})$ is obtained by multiplying two
Galilean matrices. We suppress
the index ${}_{0}$ in order to simplify notation,
and distinguish different group elements by primes (${}'$). Then
\bg
 \left( {\matrix{
1&0&v''&x''\cr
0&1&0&v''\cr
0&0&1&t''\cr
0&0&0&1 \cr} }  \right)
=
 \left( {\matrix{
1&0&v'&x'\cr
0&1&0&v'\cr
0&0&1&t'\cr
0&0&0&1 \cr} }  \right)
 \left( {\matrix{
1&0&v&x\cr
0&1&0&v\cr
0&0&1&t\cr
0&0&0&1 \cr} }  \right)
\label{matform}
\ed
implies the following composition law for $G$:
\bg
\label{galcomp}
x''= & x'+x+v't\qquad \qquad  \qquad \qquad \qquad \qquad  \cr
v''=& v'+v \qquad \qquad \qquad \qquad \qquad \qquad \qquad \cr
t''=& t'+t \qquad \qquad \qquad \qquad \qquad \qquad \qquad
\ed

\section{Lagrangian semi-invariance}
Since the equations of motion (\ref{gal1}) are invariant under $G$,
we expect that the free particle will be described by a Galilean
invariant Lagrangian ${\L}(x,v,t) $. Let us check this assumption.

Under a Galilean transformation (\ref{gal2}), the Lagrangian changes to
\bg
{\L}(x,v,t) \ra {\L}(x+x_{0}+v_{0} t,v+v_{0},t+t_{0}) \equiv
{\L}(x',v',t')
\ed
It is enough to demand invariance
under infinitesimal Galilean transformations. Then
$(x_{0},v_{0},t_{0})$ are very small and one can
expand in a Taylor series
\bg
{\L}(x',v',t')
\simeq {\L}(x,v,t) +(x_{0} \Xxr +t_{0}\Xtr +
v_{0} \Xvr ) {\L}(x,v,t) + \ldots
\ed
where
\bg
\Xxr={\D  \over \D x}  \qquad
\Xtr={\D \over \D t} \qquad
\Xvr= t {\D  \over \D x}  + {\D \over \D v}
\label{galgen1}
\ed
are infinitesimal Galilean
operators.
Invariance under  these transformations gives
\bg
\Xxr {\L}= 0 & \ra \quad  {\L}(x,v,t)={\L}(v,t)
\nonumber \\
\Xtr {\L}= 0 & \ra  \quad {\L}(x,v,t)={\L}(v)
\nonumber \\
\Xvr{\L}= 0  & \ra \quad \displaystyle{  {\D {\L}\over \D v}=0}
\ed
The most general Galilean invariant Lagrangian is a constant!~.

The solution of this apparent puzzle is the following:
since the  equations of motion derived from a
Lagrangian $L $ do not change if we add the time
derivative of an arbitrary function $F(x,t)$ \cite{f1b} ,
strict invariance
is a too stringent condition.
The Lagrangians $L $ and $L + dF/dt$
must be considered equivalent {\cite{levy}}.
We only need \cite{cursi}
semi-invariant Lagrangians that change in a total derivative
under an infinitesimal action of a symmetry group.

Define functions
$f^{(v)},f^{(t)}$ and $f^{(x)}$ of $(x,t)$, such that
\bg
\Xvr {\L} ={ d f^{(v)} \over d t} \qquad , \qquad
\Xtr {\L} ={ d f^{(t)} \over d t}\qquad , \qquad
\Xxr {\L} ={ d f^{(x)} \over d t}
\label{inv1}
\ed
Semi-invariance under the infinitesimal velocity generator, $\Xvr$ ,
implies
\bg
t {\D{\L}  \over \D x}  + {\D {\L}\over \D v}=
v {\D f^{(v)} \over \D x}  +{\D f^{(v)} \over \D t}
\nonumber \\\
\ra {\L}(x,v,t)=\a (x,t)v^{2}+\b (x,t) v + \g (x,t)
\label{s1}
\ed
where $\a (x,t),\b (x,t),\g (x,t)$ are functions to be determined.
Semi-invariance under time translations further restricts ${\L}$
\bg
\Xtr {\L} =  \displaystyle{ { d f^{(t)} \over d t} } &
\nonumber \\
\ra \quad {\D{\L}  \over \D t} = \disp{ v^2  {\D \a \over \D t}}\, \,\,+ &
\displaystyle{ v {\D \b \over \D t}+ {\D \g \over \D t}
= v {\D f^{t} \over \D x}  +{\D f^{t} \over \D t}}
\nonumber \\
\nonumber \\
\ra \quad {\D \a \over \D t} = 0 & \ra \qquad \a (x,t)=\a (x)
\label{s2}
\ed
and proceeding in the same way
\bg
\Xxr {\L} ={ d f^{(x)} \over d t} \qquad \ra  \qquad
{\D \a \over \D x} =0 \qquad \ra \qquad \a (x,t)={\rm{constant}}
\equiv {1 \over 2} m
\label{s3}
\ed
The most general Lagrangian semi-invariant under $G$ is then
\bg
{\L}(x,v,t)={1 \over 2} m v^2 +\b v + \g
\label{lag0}
\ed
where we have chosen the constant to be $(1 / 2) m $ so   that
the usual Lagrangian $~L=(1/2) m v^2~$ is obtained, plus
some additional
terms. According to our previous discussion
on semi-invariance, we expect these terms to be a total derivative.
{}From (\ref{lag0}) and (\ref{inv1})
\bg
{\D f^{(t)} \over \D t} = {\D \g \over \D t} \qquad \qquad
{\D f^{(t)} \over \D x} = {\D \b \over \D t} \nonumber \\
{\D f^{(x)} \over \D t} = {\D \g \over \D x} \qquad \qquad
{\D f^{(x)} \over \D x} = {\D \b \over \D x}
\label{int1}
\ed
Integrability of equations (\ref{int1}) implies
\bg
\b(x,t)= {\D F(x,t) \over \D x} \qquad  \qquad
\g= {\D F(x,t) \over \D t} \qquad  \qquad
\label{def1}
\ed
for some arbitrary differentiable function $F(x,t)$.
The semi-invariant Lagrangian  $\L$ is then
\bg
{\L}(x,v,t)={1 \over 2} m v^2 + v {\D F \over \D x}  +
{\D F \over \D t}  \equiv
{1 \over 2} m v^2 + {d F \over d t}
\label{lagtext}
\ed
As we expected. However, there is something important
to be learned from this well known result
(in fact, this is our justification for our
long calculations):
because semi-invariant Lagrangians are defined
up to a total derivative, equivalence classes of Lagrangians,
not individual Lagrangians, are the relevant quantities in classical
mechanics.

\section{A new symmetry}

We can learn more about the connection between symmetry
and Lagrangian semi-invariance by studying
how the arbitrary function $F(x,t)$ in (\ref{lagtext}) transforms
under $G$.

It is straightforward to verify from (\ref{inv1}), (\ref{def1})
and (\ref{lagtext}) that
\bg
f^{(t)}={\D F \over \D t} \qquad , \qquad
f^{(x)}={\D F \over \D x} \qquad, \qquad
f^{(v)}=t {\D F \over \D x} + m x = t {\D F \over \D x} +
{\D F \over \D v} + m x
\label{trans1}
\ed
We can rewrite the previous expressions as (see (\ref{galgen1}))
\bg
f^{(t)}=\Xtr \, F  \qquad ,\qquad
f^{(x)}=\Xxr \, F  \qquad ,\qquad
f^{(v)}=\Xvr \,F + m x
\label{trans2}
\ed
The semi-invariant functions $f^{(v)},f^{(t)}$
and $f^{(x)}$ can be expressed
as infinitesimal Galilean transformations acting on an arbitrary function
$F(x,t)$,
{\em{except}} for an extra linear piece $ m x $ that appears
in~$f^{v}$.

The crucial observation
is that, with the introduction of a new real variable $\th$,
(\ref{trans2}) can be written as a set infinitesimal
variations acting on  $F$. Consider functions of the form
(see (\ref{condit}))
\bg
F(x,t,\th)=F(x,t)+\th
\label{pol}
\ed
and define the following vector fields
\bg
\tec={\D \over \D \th}  \quad , \qquad
\ET  ={\D \over \D t}  \quad , \qquad
\XT  ={\D  \over \D x}  \quad ,\qquad
\KT  =t {\D  \over \D x}  + {\D \over \D v} + m x  \tec
\label{gen3}
\ed
Equations (\ref{trans1}) read now
\bg
f^{(t)}= \ET F \qquad , \qquad
f^{(x)}= \XT F \qquad , \qquad
f^{(v)}= \KT F
\label{trans3}
\ed
Moreover, these
transformations form an algebra ($[,]$ is the Lie bracket)
\bg
[\ET,\XT] = 0 \quad , \quad [\ET,\KT] = \XT \quad , \quad
[\XT , \KT] = m \tec
\label{alg1}
\ed
with $\tec$ commuting  with every generator. In contrast, the Galilean
commutators are
\bg
[\Xtr,\Xxr] = 0 \quad , \quad [\Xtr,\Xvr] = \Xxr \quad , \quad
[\Xxr , \Xvr] = 0
\label{alg2}
\ed
The relations (\ref{alg1}) indicate the existence of a new symmetry
group for the free
classical particle. Since
this symmetry does not involve the Galilean variables $(x,v,t)$
and is a consequence of
the equivalence of the Lagrangians $L$ and $L + dF/dt$, we expect
its implications to be very general,
far beyond the simple system we are considering here.

\section{Central Extension of the Galileo Group}

The new commutators (\ref{alg1}) are related to
(\ref{alg2}) by
a mathematical structure of crucial importance in physics, the
{\em{central extension}} of a group.

Consider two groups $G_{1}$ and $G_{2}$ . One can obtain
a larger group by taking the cartesian product of $G_{1}$ and $G_{2}$:
if $a,b \in G_{1}$ and $y,z \in G_{2}$, one defines a new group
with composition law $\otimes$
\bg
(a,y)\otimes(b,z) = (a \star b,y \diamond z)
\ed
where ${ \star }$ and ${\diamond}$ are the composition laws
of $G_1$ and $G_2$.
This is called the direct product
of $G_{1}$ and $G_{2}$. There is no new information in a direct product.
In physics, one encounters direct products of groups
in the study of  non interacting systems.

A more interesting way of combining the two groups is
\bg
(a,y) {\tilde{\otimes}} (b,z) = (a \star b,y \diamond z \diamond \e(a,b))
\label{semid}
\ed
where $\e : G_{1} \mapsto G_{2} $ is a function
whose properties will be described later.
The composition law (\ref{semid}) corresponds to
the  {\em{semi-direct product}}, or extension, of $G_{1}$
by $G_{2}$. Notice that $G_{1}$ is not a subgroup of $G_{1}
{\tilde{\otimes}} G_{2}$, but $G_{2}$ is always an invariant subgroup.
If $G_{2}$ is abelian this is called a {\em{central extension}}.

The commutation relations (\ref{alg1}) show that
this is just the structure obtained after enlarging the
Galileo group with the new variable $\th$:
the generators (\ref{gen3})
can be derived from the Galilean composition law
(\ref{galcomp}) and an additional line
of the form
\bg
\th''=\th'+\th + \e(x',v',t';x,v,t)
\label{ext1}
\ed
Because $\tec$ commutes with all generators, the extension
must be central. Since $\th$ is real and has an additive composition
law, this is called a central extension of $G$ by
the real numbers $R$, $G_{R}$.

The function $\e $ in (\ref{semid}) and (\ref{ext1})
is called a {\em{cocycle}}.
Cocycles are restricted by the group law properties:
associativity and existence of an inverse \cite{bargmann}.
Let $a, b ,c
\in G_{1}$. Associativity of the group law
$$ a \star (b \star c) = (a \star b) \star c $$
implies  that $\e$ must satisfy the following functional
equation
\bg
\e(a,b) \diamond \e(a \star b, c)=\e(a,b \star c) \diamond \e(b, c)
\label{coc}
\ed
Consider now an arbitrary function $f(a): G_{1} \mapsto G_{2}$,
and define \cite{galcob}
\bg
\delta_c f(a,b)= f(a \star b)\diamond (f(a))^{-1} \diamond (f(b))^{-1}
\label{borde}
\ed
One can check that $\delta_c f$ fulfills (\ref{coc}). But $\delta_c f$
amounts to a simple  change of coordinates in $ G_{2}$
$$z \rat z \diamond f(a) \qquad z \in G_{2} $$
These special cocycles, called {\em{coboundaries}}, do not define any true
extension, but rather a direct product \cite{perico}.
They are equivalent to the trivial cocycle
$\e (a,b) = 0 $.

We define an extension $G_{R}$ by the cocycle \cite{bargmann}
\bg
\e(x',v',t';x,v,t)= m ( v' x +{ 1 \over 2} t {v'}^{2} )
\label{cocycle}
\ed
The composition law  for  $G_{R}$ is then
\bg
x''= & x'+ x \, +  &  v't
\nonumber \\
v''=& v'+v &
\nonumber \\
t''=& t'+t &
\nonumber \\
\th''=& \th'+\th  \,+ & m \, ( v' x   + { 1 \over 2} t {v'}^{2} )
\label{galreal}
\ed
One obtains the expression
(\ref{gen3}) for the right-invariant generators of the group.
The left-invariant generators \cite{cohomp} correspond to variation
with  respect to unprimed parameters in (\ref{galreal})
\bg
{X_{\th}}^{L}\equiv\  \tec={\D  \over \D \th}  \quad , \qquad
\XTl= {\D  \over \D x}+ m v {\tec} \quad , \qquad
\ETl= {\D \over \D t} + v {\D \over \D x} +
{1 \over 2} m v^2 \tec \quad , \qquad
\KTl= {\D \over \D v}
\label{genlr}
\ed
If we have a particular central extension
characterized by a cocycle $\e$, and we perform a change of coordinates
in the group, we obtain the new cocycle $\e'= \e + \delta_c f$, where
$\d f$ is the coboundary (\ref{borde})
associated to the change of coordinates.
This means that
two cocycles $\e$ and $\e'$ must be considered equivalent if
they differ in a coboundary.
For instance, in his study of the Galileo group,
Bargmann \cite{bargmann} considered the cocycle
\bg
\e_{B}(x',v',t';x,v,t)= - {1 \over 2} m \, ( x' v - v' x +v v' t)
\label{bargmann}
\ed
which can be obtained from (\ref{cocycle})
by adding the coboundary
generated by
\bg
f(x,v,t)=- {1 \over 2} m  \, x v
\ed
This equivalence relation defines an
important concept, the {\em{cohomology}} of a group
\cite{bargmann}.

Use of (\ref{bargmann}) instead of (\ref{cocycle})
in the extended group law (\ref{galreal})
would change the expression for the $\tec$ dependent part
of the generators. This should not affect
the dynamical description of the free particle, because
adding a coboundary to the group law
amounts to replace a Lagrangian
$L$ for $L + dF/dt$.

It is important to realize that
the cocycle ${\bar{\e}} = {\bar{m}} ( v' x +{ 1 \over 2} t {v'}^{2} )$,
with ${\bar{m}} \neq  m$,
is not equivalent to $\e$, since $\e - {\bar{\e}}$
is not a coboundary: the mass of the particle has a cohomological
significance,  it parametrizes the extensions of the Galileo group.

\section{Geometrical structure}
\subsection{Fiber Bundle structure. The canonical form $\Theta$}

Our central extension $G_{R}$ singularizes the transformations
generated by $\tec$:
two points $a,b\in G_{R}$ that differ in the value of the
$\th$ variable are equivalent and give the same description
of the free particle,
since $\th$ is not a measurable quantity, like position or velocity.
In some sense, $\th$
is irrelevant. Nevertheless, as in the case for gauge
invariance in electromagnetism, the requirement that $\th$ must
not be observable has a profound dynamical content  and is responsible
for a geometrical structure, a fiber bundle,
which is  ubiquitous in many areas of physics.

First, note that although
$G$ is not a subgroup of $G_{R}$, $G_{R}/R \simeq G$ as
topological spaces:
any $g \in G_{R}$ can be decomposed in two parts, $g=(a,\th)$, where
$a \equiv (x,v,t) \in G$ and $\th \in R$. This defines a projection
$\pi: G_{R} \mapsto G  \quad / \, \pi(x,v,t,\th)=(x,v,t)$.

The triplet $ (G_{R}, R, \pi) $
is a {\em{fiber bundle}} (\cite{isham}, \cite{schutz}) :
$G_{R}/R $ is called {\em{base}} manifold, $R$ is the
{\em{fiber }} and $\pi$ is the projection from $G_{R}$ to the base
manifold. Since the fiber $R$ has a
group structure, we say that
$(G_{R},R,\pi)$ is a fiber bundle with structural group $R$.

One can visualize such geometric object by imagining that through
each point in the base manifold there is a line (fiber) of points
with different values of the fiber coordinate $\theta$.
It is natural then to call {\it{vertical}} any transformation or
quantity that involves only $\th$,
whereas quantities defined on the base manifold $G$ are called
{\it{horizontal}}.

Since we have a group
structure, it is possible to define verticality or horizontality
in an invariant way. Consider \cite{aldaya}
the dual form of $\tec$
\bg
\Theta(\tilde{X^{L}}) =0  \qquad \qquad \Theta(\tec)=1
\label{fff}
\ed
where $\tilde{X^{L}}$ stands for all left-invariant generators  other than
$\tec$.
$\Theta$ is the {\em{vertical part of the canonical left form}}
(\cite{isham}, p. 83). We will call $\Theta $
just {\em{canonical form}}, for brevity.

Geometrically, $\Theta$ is a connection (\cite{isham},
\cite{schutz}) in the fiber bundle, and
its differential $\omega = d \Theta$ is a curvature form.
A quick calculation gives
\bg
\Theta= -m v \, d x  + {1 \over 2} m v ^{2} \, d t+ d \th
\label{form1}
\ed
Why do we  consider such an object~?
We give an intuitive argument: $\Theta$ is blind to the
physical part (the Galileo group, the horizontal direction ) of the bundle,
that is, $\Theta$ is zero on any linear combination of (left) pure
Galilean
generators. Symmetries of $\Theta$, in a sense that will be made clear
later, must be or must originate physical quantities .
This is substantiated
by the following: restrict $\Theta$ to the base manifold $G$
\bg
\Theta_{\vert_{G}} = -m v \, d x  + {1 \over 2} m v ^{2} \, d t
\ed
Rewrite the above in the form
\bg
\Theta_{\vert_{G}} = -m v \, (d x - v \,d t)
  - {1 \over 2} m v ^{2} \, d t
\ed
Then, along trajectories such that $dx = v d t\quad \ra v = dx / dt $
\bg
\Theta_{\vert_{G}} =  - {\cal{L}} d t
\ed
where $\cal{L}$ is the free Lagrangian. Hence, the quantity
\bg
S = \int{ \,\Theta_{\vert_{G}} }
\label{action1}
\ed
is analogous to the classical action, or more precisely,
to the Poincar\'e-Cartan form of analytical mechanics
\cite{symp}.

\subsection{Symmetries}

A transformation $X$ is a symmetry of $\Theta$ if
$X$ leaves $\Theta$ semi-invariant,
that is, the {\em{Lie derivative}} (see appendix I) of $\Theta$
with respect to the considered
transformation $X$ is a total differential
\bg
L_{X} \, \Theta = d f^{X} \qquad \ra \qquad
d (i_{X} \Theta) + i_{X} (d \Theta)  = d f^{X}
\label{lie}
\ed
where $f^{X}$ is some function associated with $X$. From (\ref{form1})
\bg
\omega = d \Theta = m d x \wedge d v + m v d v \wedge d t
\label{omega}
\ed
The characteristic module of $\Theta$ ,
$~{\cal{C}}_{\Theta}= {\rm{ker}}\, \Theta \cap {\rm{ker}}\, d \Theta$ is
the set of
transformations that leave
$\Theta $ {\em{strictly}} invariant.
\bg
d (i_{X} \Theta) = 0 \quad, \quad  i_{X} (d \Theta)  = 0
\qquad \forall X \in {\cal{C}}_{\Theta}
\label{modul}
\ed
It is straightforward to check that the time translation generator
$\ETl$ is the only generator of ${\cal{C}}_{\Theta}$. Its
integral curves (appendix I)  give the equations of motion for the
free particle.
\bg
\Xtl=
{\D \over \D t}+ v{\D \over \D x}+ {1 \over 2} m v^{2}\tec
\qquad  \ra \qquad {d t \over d s} =1 \quad ,  \quad
{d x \over d s} =v   \quad , \quad
{d v \over d s} = 0   \quad , \quad
{d \th \over d s} = {1 \over 2} m v^{2}
\label{intcur}
\ed
where $s$ is a parameter. Then
\bg
x(s)= x_{0}+ v_{0} s \quad , \quad
v(s)= v_{0} \equiv {p \over m} \quad , \quad
t(s) = s \quad ,\quad \th(s) = E s
\label{curves}
\ed
where $E \equiv (1/2) m {v_{0}}^2 = p^2/2 m$.
In appendix B,
we show how
the characteristic module defines the classical
Poisson bracket structure (\cite{aldaya}, \cite{symp} ,
\cite{rigormortis}).

While the free particle Lagrangian is only semi-invariant
under Galilean transformations,
$\Theta$ is invariant  under
the right generators (\ref{galgen1}) of $G_{R}$:
$L_{X^{R}} \Theta = 0$ with $X^{R}$ any right generator of $G_{R}$.
This is the basis \cite{rigormortis}
of the group formulation of Noether's theorem : we prove in
appendix II that
the functions $h^{X^{R}} = \Theta (X^{R})$ are constants of motion
(constant on the trajectories  (\ref{curves})). For instance
\bg
\Theta(\XT)= (-m v \, d x  + {1 \over 2} m v ^{2} \, d t+ d \th)
({\D \over \D x})= -m v ({\D x \over \D x})
+ {1 \over 2} m v ^{2} ({\D t \over \D x}) + ({\D \th \over \D x}) = -m v
\equiv - p
\label{no1}
\ed
Also
\bg
\Theta(\ET)= - E \qquad
\Theta(\KT)= m (x-v t)  = m x_{0}
\label{no2}
\ed
As it should, translation invariance implies momentum conservation and
time invariance is responsible for energy conservation.
We find also that {\it{velocity}}
invariance results in the conservation of the initial position
$x_{0}$.

Now, we build invariant functional spaces.
Define, analogously as in (\ref{pol})
\bg
{\cal{F}} = \{ \Psi : G_{R} \mapsto R  \quad / \quad
\Psi (\th + g) = \Psi(g)+ \th \qquad \forall g \in G_{R}
\label{condit}
\ed
In coordinates
\bg
{\cal{F}} = \{ \Psi : G_{R} \mapsto R  \quad
/ \quad \Psi(x,v,t,\th)= \Psi(x,v,t)+\th \}
\label{polgal0}
\ed
The definition (\ref{condit})
reflects the triviality of the $\th$ dependence for physical
quantities. Notice that $\tec \, \Psi = 1 \quad
\forall \Psi \in {\cal{F}}$.
We further constrain ${\cal{F}}$
by imposing invariance under $G_{R}$
transformations. We would like ${\cal{F}}$ to be invariant under the generator
of ${\cal{C}}_{\Theta}$, since its integral curves define the
equations of motion. What others constraints can we impose ?
If one wants a maximal set of constraints on $\Psi$
(this is called a polarization in the Geometric Quantization formalism),
compatibility is the only requirement: it is not possible
to have invariance under transformations (differential
operators) that do not close an algebra.
The generator of the characteristic module is $\ETl$, and
$\tec$ commutes with all generators. By (\ref{alg1}), we see
that
\bg
{\cal{P}} = \{ \ETl,\XTl \}
\label{polgal}
\ed
is the only acceptable possibility: if we include $\KTl$, then its
commutator  with $\XTl$ will be proportional to $\tec$,
and $\tec \Psi = 0 $ is incompatible
with the definition of ${\cal{F}}$.
${\cal{P}}$ is called the
polarization subalgebra, and
\bg
{\cal{F}}_{\cal{P}} =
\{ \Psi \in {\cal{F}} \quad / \quad X \Psi = 0
\quad \forall X \in {\cal{P}} \}
\label{polfundef}
\ed
is the space of polarized
functions \cite{perel}.

We have
\bg
\ETl \Psi= 0 \quad \ra
{\D \Psi \over \D t}+ v{\D \Psi\over \D x}+ {1 \over 2} m v^{2}=0
\qquad \ra \quad
\Psi(x,v,t,\th)=-m v x - S(v,t)  \nonumber
\ed
\bg
\XTl \Psi = 0 \quad \ra \quad {\D S \over \D t}+{p^2 \over 2 m }=0
\label{ham-jac}
\ed
where $p = m v$.
We recognize the Hamilton-Jacobi equation in
momentum space (\cite{rigormortis},\cite{goldstein})
\bg
{\D S \over \D t}+H (p, -{\D S \over \D p}) = 0
\label{hamijac}
\ed
The term in ${\D S / \D p}$ is absent due to the simple form of
the  Hamiltonian $H(p,x) = H(p) = p^2 / 2 m $

\section{Quantum mechanics}

Classical and quantum mechanics
are beautifully related in our group invariance formalism
(\cite{aldaya}, \cite{bargmann}). First,
we note that the
cocycle $\e$ in (\ref{ext1}), and hence $\th$,
has dimensions of action. Let's introduce $\hbar $ explicitly
\bg
\th''=\th'+\th + {1 \over \hbar} m ( v' x +
{ 1 \over 2} t {v'}^{2} )
\ed
$\th$ is now dimensionless. We can consider the above line as the
uncompactification of a U(1) multiplicative law
\bg
\zeta \equiv e^{-i \th} \in U(1)
\label{circle}
\ed
and obtain an extension of $G$ by U(1), $\tilde{G}$
\bg
x''= & x'\, \, + & x  +  v't
\nonumber \\
v''=& v'\, \, + & v
\nonumber \\
t''=& t'\, \, + & t
\nonumber \\
\zeta''=& \zeta' \zeta & \exp \quad ( -{  i  \over \hbar} m ( v' x +
{ 1 \over 2} t {v'}^{2} ) )
\label{ext2}
\ed
Note that $\th$ in (\ref{circle}) is not  restricted to the interval
$[0, 2\pi )$.

We can apply now the formalism developed in the last sections in order
to study the dynamics defined by $\tilde{G}$.
The expressions for the generators and canonical form $\Theta$ are
\bg
\XTl={\D  \over \D x}- {1 \over \hbar} m v \Xi \qquad
\ETl={\D \over \D t} + v {\D \over \D x}
-{1 \over 2 \hbar} m v^2 \Xi \qquad
\KTl={\D \over \D v} \qquad \
{X_{\zeta}}^{L}\equiv\Xi=i \zeta {\D  \over \D \zeta}
\ed
\bg
\XT={\D  \over \D x} \qquad \qquad
\ET={\D \over \D t} \qquad \qquad
\KT= t {\D  \over \D x}  + {\D \over \D v}-
{1 \over \hbar}m x \Xi \qquad \qquad
{X_{\zeta}}^{R}\equiv\Xi=i \zeta {\D  \over \D \zeta}
\ed
\bg
\Theta= -m v d x + {1 \over 2} m v ^{2} d t -  {i \over \zeta} d \zeta
\ed

As in the case of $G_{R}$, ${\cal{C}}_{\Theta}$
is generated by the (left) time translations. The quantum
equations of motion are the integral curves
of $\ETl$
\bg
{d t \over d s} = 1 \quad , \quad
{d v  \over d s} = 0 \quad , \quad
{d x  \over d s} = v \quad , \quad
{d \zeta \over d s} = {i \over 2 \hbar} \,\zeta
\ed
\bg
\ra
t = s \quad , \quad v = p/m \quad ,
\quad x = x_{0} + {p \over m}  t \quad , \quad
\zeta = \zeta_{0} \, \exp ({i \over \hbar} E t) \qquad E =
{p^{2} \over 2 m }
\label{mod1}
\ed
The polarization subalgebra is
\bg
{\cal{P}} = \{ \ETl,\XTl \}
\label{polu1}
\ed
formally the same as (\ref{polgal}).
In analogy with (\ref{condit}) and (\ref{polfundef}),
we consider complex polarized
functions
\bg
{\cal{F}}_{\cal{P}} =
\{ \Psi(x,v,t,\zeta) : \tilde{G} \mapsto {\bf{C}} \quad
/ \quad \quad \Psi(x,v,t,\zeta)= \zeta \, \Psi(x,v,t)
 \quad , \quad X \Psi = 0
\quad \forall X \in {\cal{P}} \}
\ed
Note that $\Xi \Psi = i \Psi $

Invariance under ${\cal{P}}$ gives the following equations for $\Psi$
\bg
\XTl \Psi = 0 \quad \ra {\D \over \D x} \Psi - {i \over \hbar}
p \Psi = 0
\label{sch0}
\ed
\bg
\ETl \Psi = 0 \quad \ra \quad {\D  \over \D t} \Psi +
{p \over m} {\D  \over \D x} \Psi - {i \over \hbar} {p^2 \over 2 m} \Psi =0
\label{sch1}
\ed
The identification of the
momentum operator
\bg
\hat{p} \equiv - i \hbar {\D \over \D x}
\label{polop}
\ed
is straightforward from (\ref{sch0}).
Combining (\ref{sch0}) and (\ref{sch1}), we obtain
\bg
i \hbar {\D  \over \D t} \Psi ={p^2 \over 2 m} \Psi \qquad
\ra \qquad
i \hbar {\D  \over \D t} \, \Psi ={\hat{H}} \, \Psi \quad ,
\qquad {\hat{H}} = {{\hat{p}^2} \over 2 m}
\label{scho3}
\ed
which is the Schr\"odinger equation for the
free quantum particle wave function $\Psi$ in the momentum representation.
This is to be compared with
equation (\ref{ham-jac}), the classical Hamilton-Jacobi equation.

A basis for the solutions of (\ref{scho3}) are plane waves modulated
by an amplitude $\psi(p)$
\bg
\Psi = \zeta  \,  \exp(-{i \over \hbar}(E t -p x))   \, \psi(p)
\label{wave}
\ed
Commutativity of the right and left generators makes
the right invariant fields good candidates for
quantum operators:
because ${\cal{P}}$ is spanned by left invariant fields, if
$ X^{R} $ is a right generator and $\Psi$ is polarized,
then
\bg
X^{P} \Psi' = X^{P} (X^{R} \Psi) = X^{R} (X^{P} \Psi) = 0
\quad \quad \forall X^{P} \in {\cal{P}} , \quad \forall \Psi \in
{\cal{F}}_{\cal{P}}
\ed
so $\Psi'$ is polarized and
it has the general form (\ref{wave}). Hence, the action of the right
invariant generators on ${\cal{F}}_{\cal{P}}$
is well defined. Note that the momentum operator in (\ref{polop})
can be written as
\bg
\hat{p}= -i \hbar \XT
\ed
Consider the action of $\hat{p}$ and
\bg
\hat{x} \equiv  { i \hbar \over m} \KT
\ed
on the amplitudes $\psi(p)$
\bg
\hat{p}: \psi \mapsto \psi'
\quad / \quad \hat{p} (\Psi) =
\zeta \, \exp(-{i \over \hbar}(E t -p x)) \, \psi'(p)
\ed
and analogously for $\hat{x}$.
After a short calculation, one obtains
\bg
\hat{p}: \psi(p) \mapsto p \, \psi(p) , \qquad
\hat{x}: \psi(p) \mapsto i \hbar {\partial \over \partial p} \, \psi(p)
\ed
These are the quantum momentum and position operators in the momentum
representation, with the Heisenberg commutation relations
\bg
[\hat{x},\hat{p}] = i \hbar
\ed
If  the functions $\psi$ are square integrable, one can
define the scalar product in the usual
way
\bg
\langle \Psi_{1} \vert \Psi_{2} \rangle =
\langle \psi_{1} \vert \psi_{2} \rangle = \int{\,d p \,
{\psi^{*}_{1}(p)} \,\psi_{2}(p) }
\label{scalprod}
\ed
Note
that the measure $d p$ is (proportional to) the dual form of the
generator not present in the polarization ${\cal{P}}$, $\KTl$.

This ends our analysis of the quantum mechanics of the free
particle, given by the extension of the Galileo group by $U(1)$.
Since classical mechanics
is described by the extension of $G$ by $R$, $G_{R}$, there is,
as promised,
a beautiful characterization for the classical limit of a quantum
theory in this group formalism: the classical theory
is obtained by opening the $U(1)$ (a circle) in the group extension
to $R$ (a line).

\section{Conclusions}

We have constructed, step by step, a group theoretical formalism
that can describe physical systems possessing a dynamical group
with a fiber bundle structure. We found that for the particular examples
of the Galileo extensions ${\tilde{G}}$ and $G_{R}$ one obtains the
quantum and classical mechanics of a free particle, with dynamical
quantities like Noether invariants, operators and  evolution
equations obtained from geometrical objects canonically defined
in the group.

Here are the more relevant features of this formalism
\begin{itemize}
\item{}{The physical system is described by a central extension
        of a Lie group, which has a natural fiber bundle structure.}
\item{}{Noether invariants are given by symmetries of
        the canonical left-form $\Theta$, dual to the extension parameter.
        $\int \Theta$ is analogous to the action in classical mechanics.}
\item{}{The characteristic module of $\Theta$,
         ${\cal{C}}_{\Theta}$, determines the equations
        of motion. The symplectic structure (Poisson brackets) is
        obtained when $d \Theta $ is restricted to ${\cal{C}}_{\Theta}$.}
\item{}{Evolution equations and Hilbert space of states are defined
        by a (horizontal) maximal set of constraints,
        the polarization subalgebra,
        built from the left-invariant generators.}
\item{}{The right invariant generators are well defined operators in
        the space of polarized functions.}
\end{itemize}.

Although we have checked the points above in
our simple example, we have not proven them generally. Rigorous
proofs can be found in \cite{aldaya} and \cite{rigormortis}.

Obviously, not every conceivable physical system has this
geometric structure, or even if that were true, finding
the corresponding dynamical group for an arbitrary
system would be extremely difficult. But since symmetry plays such a
crucial role  in the foundations of physics, we believe
that it is of interest
to see how much information can be extracted from symmetry
considerations alone. And with the added bonus of aesthetic
beauty.

\vskip 2 cm
{\bf{Acknowledgments}} I am indebted to J.A. Azc\'arraga
and V. Aldaya for having taught me everything I know about Quantization on
a Group. G. S. Guralnik,
M. Espinosa, A. Jevicki and S. Hahn gave me useful comments.  This
work has been supported in part by DOE Grant DE-FG02-91ER40688 - Task D
and NSF Grant ASC-9211072.

\newpage
\section{Appendix I}
\subsection{Vector fields}
In this appendix we give  concise
definitions of some geometric objects, oriented towards
explicit calculations so that the reader could follow the
computations in the main text.
More comprehensive definitions can be found in introductory
books on differential geometry (\cite{isham}, \cite{schutz}).

Let $M$ be a N-dimensional manifold with local
coordinates $x_{i} \quad i =1,2 \ldots N $. A {\em{vector field }}
$X$ is an application that associates a first order
differential operator $X({\bf{x}})$ to a point ${\bf{x}} \in M$.
$X({\bf{x}}) $ can be expressed in local coordinates as a
linear combination of the base fields
\bg
e_{i} \equiv {\D \over \D x_{i}}
\ed
That is
\bg
X({\bf{x}}) = \sum_{i=1}^{N} X_{i}({\bf{x}})
{\D \over \D x_{i}}
\ed
with $X_{i}({\bf{x}}) \quad i=1,2 \ldots N $,
differentiable functions on $M$.
The space of the vector fields is called the {\em{tangent}} space
of $M$, $T(M)$.
It is cumbersome to write $X({\bf{x}})$,
so we will write just $X$.

The {\em{integral curves}} of $X$ are the solution to the
set of ordinary differential equations
$$ {d x_{i} \over d s} = X_{i}({\bf{x}}(s)) \qquad i=1,2 \ldots N $$
where s is an integration parameter. For instance, in a $N=2$ dimensional
manifold, the integral curves for
the vector field
\bg
 Y= x_{2}{\D \over \D x_{1}} - x_{1}{\D \over \D x_{2}}
\ed
are  given by the differential equations
$${d x_{1} \over d s}= x_{2} \ra {d x_{2} \over d s}= -x_{1} $$
\bg
 \ra \quad x_{1}(s)=a \sin(s) + b \cos(s) \qquad
x_{2}(s)=a \cos(s) - b \sin(s)
\label{ccp}
\ed
with a and b constants.
Note that
the invariance condition $X f = 0$ implies that
$f$ is constant along the integral curves of $X$. In our case,
the general solution of $Y f = 0 $ is
$f=f(x_{1}^2+x_{2}^2) = f(a^2+b^2)={\rm{constant}}$.

\subsection{forms}

A 1-form $\Gamma$ is an application that associates
to every point ${\bf{x}} \in M$
an element of the dual space of $T(M)$.
\bg
\Gamma: x \mapsto \Gamma({\bf{x}}) \qquad \qquad  / \quad
\Gamma({\bf{x}}) (X({\bf{x}})) = f({\bf{x}})
\ed
with $f({\bf{x}})$ a differentiable function. As with vector fields,
we write $\Gamma$ for $\Gamma({\bf{x}})$.
The space of 1-forms is called
the cotangent space of $M$ and is denoted by $T^{*}(M)$.

A convenient representation for the basis of $T^{*}(M)$ is
\bg
u_{j} \equiv d x_{j} \qquad, \qquad j=1,2\ldots N
\ed
and its action on the basis of $T(M)$,
$e_{i} = \partial / \partial x_{i}$
\bg
u_{j}(e_{i}) \equiv d x_{j} ({\D \over \D x_{i}}) \equiv
{\D x_{j} \over \D x_{i}} = \d_{i,j}
\ed
2-forms, 3-forms , etc. are linear combinations of
tensor products ($\otimes$) of $u_{j}$. A
function $f({\bf{x}})$ is considered a zero-form.

An important operation on forms is the differential $d$. The differential
of a n-form is a (n+1)-form (or zero). For a function $f$
$$
d f = {\partial f \over \partial x_{i}} \, d x_{i}
$$
which is the definition of the ordinary differential, For
a  1-form $\Gamma = \Gamma_{i} d x_{i}$
$$d \Gamma = {\D \Gamma_{i} \over \D x_{j}} d x_{i} \wedge d x_{j} $$
where the wedge operator $\wedge$ is the antisymmetric combination
$d x_{i} \wedge d x_{j} = d x_{i} \otimes d x_{j} -d x_{j} \otimes d x_{i} $.
One can define in an analogous way differentials of higher order forms.
Note that the antisymmetry of the wedge operator implies that $d (d \Gamma)
\equiv d^{2} \,  \Gamma = 0 $ and $d^{2}\, f = 0 $

If a n-form acts on $n-k$ vector fields one obtains a k-form. For instance
the 1-form $d f$ acting on a field $X$ gives a zero-form
$$ d f (X) = {\partial f \over \partial x_{i}} \, d x_{i}
(X_{j}(x) {\partial \over \partial x_{j}}) =
X_{i}{\partial f \over \partial x_{i}} = X (f) $$
$d \Gamma$ acting on two fields $X$ and $Y$ also gives a zero-form
$$d \Gamma( X,Y) = {\D \Gamma_{i} \over \D x_{j}} (X_{i} Y_{j}-X_{j} Y_{i})$$
$d \Gamma$ acting on $X$ alone gives a 1-form
$$d \Gamma( X,.) = {\D \Gamma_{i} \over \D x_{j}}
(X_{i} d x_{j}-X_{j} d x_{i})$$

We often use the inner product notation
$$i_{X} \Omega = \Omega(X) $$
to denote the action of a n-form $\Omega$  on a vector field $X$.

\subsection{Lie Derivative}

The Lie derivative is a generalization of the directional
derivative for both vector fields and forms.
The Lie derivative of a function $f$ with respect to a vector
field $X$ is
\bg
L_{X} \,F = X (f) = df (X)
\ed
For two vector fields $ X,Y $, the
Lie derivative of $Y$ with respect to $X$, $L_{X} Y$, is a
vector field defined by
$$L_{X} Y = [X,Y] =(X_{i}{\D Y_{j} \over \D x_{i}} -
Y_{i}{\D X_{j} \over \D x_{i}}) {\D \over \D x_{j}}
= - L_{Y} X$$
That is, the ordinary Lie commutator.

The Lie derivative of a 1-form $\Gamma$ with respect to $X$, $L_{X} \Gamma $,
is also a 1-form, and is defined to be the rate of change of $\Gamma$
along the flow lines of $X$
(\cite{isham}, p. 45 ).
One finds, in local coordinates
\bg
{L_{X} \Gamma} = ({L_{X} \Gamma})_{i} \, d x_{i} \qquad \quad /
\qquad ({L_{X} \Gamma})_{i} =
X_{j} {\partial \Gamma_{i} \over \partial x_{j}} +
\Gamma_{j} {\partial X_{j} \over \partial x_{i}}
\ed
or, in a more concise notation
\bg
L_{X} \Gamma = d \,(\Gamma(X)) + (d \Gamma)\, (X,.)
\equiv i_{X} \, d \Gamma + \, d (i_{X} \Gamma)
\ed

\newpage
\section{Appendix II}
\subsection{Poisson Brackets and Symplectic Structure.}
We show explicitly that the restriction of $d \Theta$
to the characteristic module ${\cal{C}}_{\Theta}$ provides a
symplectic structure where Poisson brackets can be naturally defined.
Notice that, since time translations generate
${\cal{C}}_{\Theta}$, this quotient procedure eliminates the
time evolution.

Consider the differential of the left-invariant form $\Theta$
\bg
\omega = d \Theta = m d x \wedge d v + m v d v \wedge d t
\label{omega2}
\ed
The restriction of $\omega$ to ${\cal{C}}_{\Theta}$
(see (\ref{curves})) is
\bg
{\omega}_{{\cal{C}}_{\Theta}} \equiv {\bar{\omega}}=
m d( x_{0} + v_{0} t) \wedge  d v_{0} + m v_{0} \,d v_{0} \wedge d t =
m d x_{0} \wedge  d v_{0}  + m v_{0} \, d t \wedge d v_{0} +
m v_{0} \,d v_{0} \wedge d t
\nonumber
\ed
\bg
\ra {\bar{\omega}}= d q \wedge d p
\ed
where we have defined $q \equiv x_{0} , \quad p \equiv m v_{0}$.
In matrix form,
\bg
{\bar{\omega}} = {\bar{\omega}}_{i,j} \, d x_{i}\otimes d x_{j} \qquad
/ \quad x_{1}= q , \quad x_{2}= p , \quad
{\bar{\omega}}_{i,j}= \left( {\matrix{
0&1\cr
-1&0 \cr}}   \right)
\label{simpex}
\ed
The matrix above is the invariant metrics for the symplectic
group (see \cite{symp} or chapter 9 of \cite{goldstein})).

The form $\bar{\omega}$
establishes a correspondence between functions
$f=f(q,p)$ and vector fields: a differential operator $X_{f}$ is
associated to a function $f$ such that
\bg
i_{X_{f}} \bar{\omega} = - d f
\label{def1f}
\ed
in other words, $X_{f}$ is a symmetry (see (\ref{lie})) of
$\bar{\omega}$.
One obtains
\bg
X_{f} = {\partial f \over \partial q} {\partial \over \partial p} -
{\partial f \over \partial p} {\partial \over \partial q}
\label{def2f}
\ed
The Poisson bracket $\{ f,g\}$ of two functions $f,g$ is given by
\bg
\{f,g\} = \bar{\omega} (X_{f},X_{g})
\label{xfxg}
\ed
In coordinates
\bg
\{f,g\} = {\partial f \over \partial q} {\partial g \over \partial p}-
{\partial f \over \partial p} {\partial g \over \partial q}
= X_{f} g = - X_{g} f
\label{ddd}
\ed

Notice that the integral curves for
(\ref{def2f}) are formally analogous to to the
Hamilton equations of motion. If $f \equiv H $ in (\ref{def2f}),
the integral curves are ($t$ is an integration parameter)
\bg
{d p \over d t} = - {\partial H \over \partial q}
\qquad , \qquad
{d q \over d t} =   {\partial H \over \partial p}
\label{hameq}
\ed
Consider a hamiltonian $H(p,q)$ and its associated $X_{H}$.
It is evident, from (\ref{ddd}) that if the symmetries $X_{f}$ of
${\bar{\omega}}$ leave the hamiltonian $H$ invariant, then
\bg
L_{X_{f}} H = X^{a} H = 0 \ra \{f,H\} = 0
\ed

Note that the correspondence (\ref{def1}) that associates to a function
$f$ the vector field defined in (\ref{def2f}) is not an isomorphism: all
constant functions are mapped to the vector field zero. Moreover,
the vector fields associated to the functions $q, \, p$
$$X_{q}={\partial \over \partial p} \qquad, \qquad
X_{p}=-{\partial \over \partial q} $$
commute, while its Poisson bracket is $\{q,p\} = 1$.
If we compare the commutators (\ref{alg1}) and (\ref{alg2}), we
see that the Lie algebra of $G_{R}$ is isomorphic to the {\em{classical}}
Poisson brackets of the free particle. This is the key for building
the quantum theory \cite{geom},\cite{aldaya}.

\subsection{Noether's Theorem}

Let ${\bar{G}}$ be a central extended group with fiber
bundle structure (section 6), and $X^{R}$
a right invariant generator.
We proof now that the functions (see (\ref{no1}) and (\ref{no2}))
\bg
f^{R}=  i_{X^{R}} \Theta
\ed
are constants of motion. This constitutes the Noether theorem
in our group theoretical formulation. An alternative proof based in the
invariance of $\Theta$ under right transformations,
$L_{X^{R}} \Theta = 0$, can be found in \cite{rigormortis}.
See also \cite{nuovo}, pp. 19-21.

The function $f^{R}$ is a constant of motion if and only if
\bg
L_{Y} f^{R} = Y (f^{R}) = 0
\quad ,\qquad \forall \, Y \in {\cal{C}}_{\Theta}
\ed
since the equations of motion are the integral curves of the generators
of ${\cal{C}}_{\Theta}$. We have
\bg
L_{Y} f^{R} = L_{Y} \,(i_{X^{R}}\, \Theta ) =
i_{X^{R}}\, (L_{Y}\, \Theta)+ i_{[Y,X^{R}]}\, \Theta
\ed
where the identity
\bg
i_{[X,Y]} = L_{X} \,i_{Y} - i_{X} \,L_{Y}
\ed
has been used.
Since $ Y \in {\cal{C}}_{\Theta}$, $L_{Y} \Theta = 0$. Then
\bg
L_{Y} f^{R} = i_{[Y,X^{R}]}\, \Theta
\ed
We prove now that $[Y,X^{R}] \in {\cal{C}}_{\Theta}$,
so the right hand
side of the above equation is zero. Let $Y = f_{i} Y_{i}$,
where $f_{i}$ are functions and $Y_{i}$ are the generators
of ${\cal{C}}_{\Theta}$ . Consider
\bg
[Y,X^{R}]  = [f_{i} Y_{i},X^{R}]  =
f_{i}[ Y_{i},X^{R}] + X^{R}(f_{i}) \,Y_{i}
\ed
$[ Y_{i},X^{R}] = 0 $
because ${\cal{C}}_{\Theta}$ is generated by left invariant fields
and left and right generators commute. Then
\bg
[Y,X^{R}] = X^{R}(f_{i}) \,Y_{i} \in {\cal{C}}_{\Theta}
\qquad \quad
\ra \qquad L_{Y} f^{R} = 0
\qquad \forall \, Y \in {\cal{C}}_{\Theta}
\ed


\begin{thebibliography}{99}
\bibitem{f1b}{This is because the
equations of motion are second order in time.
See \cite{rigormortis}
for examples of higher order equations where $F$ can also
depend on $v \equiv dx/dt$.}
\bibitem{cursi}{This is not exactly true. The quantity that
must remain invariant  is the action. Requirement of
invariance under some groups, like
the Conformal group (but not Galileo or Poincar\'e groups)
affect the volume form in the action integral.
Lagrangian semi-invariance is no longer the condition for the
variation of the action to be zero.}
\bibitem{galcob}{
In the case of
the extensions of the Galileo group
by $R$, (\ref{borde}) reads
$$\delta_{c} f(a,b)= f(a \star b)-f(a)-f(b)$$
since the group law in $R$ is additive.}
\end{thebibliography}
\end{document}